\documentclass[aps,prl,twocolumn,showpacs,groupedaddress,preprintnumbers]{revtex4}
\usepackage{graphicx}  
\usepackage{subfigure}
\usepackage{dcolumn}   
\usepackage{bm}        
\usepackage{amssymb}   
\usepackage{ulem}
\usepackage{dcolumn}
\usepackage{epsfig}

\newcommand{\gev}  {\mbox{${\rm GeV}$}}

\newcommand{\invfb}{\mbox{${\rm fb}^{-1}$}}

\newcommand{\etal} {\mbox{$et~al.$}}

\newcommand{\pt}  {\mbox{$p_{\rm T}$}}
\newcommand{\ptvis}  {\mbox{$p_{\rm T}^{\rm vis}$}}
\newcommand{\et}  {\mbox{$E_{\rm T}$}}

\newcommand{\met} {\mbox{${E\!\!\!\!/_{\rm T}}$}}

\newcommand{\mtautaupeak}{\mbox{$M_{\tau\tau}^{\rm peak}$}}

\newcommand{\mjttpeak}{\mbox{$M_{j\tau\tau}^{\rm peak}$}}

\newcommand{\mjtpeak}{\mbox{$M_{j\tau}^{\rm peak}$}}
\newcommand{\meff}{\mbox{$M_{{\rm eff}}$}}
\newcommand{\meffpeak}{\mbox{$M_{{\rm eff}}^{\rm peak}$}}

\newcommand{\meffbpeak}{\mbox{$M_{{\rm eff}}^{(b)\rm peak}$}}
\newcommand{\dM}{\mbox{$\Delta M$}}

\newcommand{\azero}{\mbox{$A_{0}$}}

\newcommand{\mzero}{\mbox{$m_{0}$}}
\newcommand{\mhalf}{\mbox{$m_{1/2}$}}

\newcommand{ \gluino}   {\mbox{$\tilde{g}$}}
\newcommand{ \squark}   {\mbox{$\tilde{q}$}}
\newcommand{ \squarkL}   {\mbox{$\tilde{q}_{L}$}}

\newcommand{ \usquarkL}  {\mbox{$\tilde{u}_{L}$}}
\newcommand{ \usquarkR}  {\mbox{$\tilde{u}_{R}$}}

\newcommand{ \sbottomone}{\mbox{$\tilde{b}_{1}$}}

\newcommand{ \sbottomtwo}{\mbox{$\tilde{b}_{2}$}}

\newcommand{ \stopone}  {\mbox{$\tilde{t}_{1}$}}

\newcommand{ \stoptwo}  {\mbox{$\tilde{t}_{2}$}}

\newcommand{ \sleptonR}  {\mbox{$\tilde{\ell}_{R}$}}
\newcommand{ \seleR}    {\mbox{$\tilde{e}_{R}$}}

\newcommand{ \seleL}    {\mbox{$\tilde{e}_{L}$}}

\newcommand{ \stauone}  {\mbox{$\tilde{\tau}_{1}$}}

\newcommand{ \stautwo}  {\mbox{$\tilde{\tau}_{2}$}}

\newcommand{ \schionezero }{\mbox{$\tilde{\chi}_{1}^{0}$}}
\newcommand{ \schitwozero }{\mbox{$\tilde{\chi}_{2}^{0}$}}

\newcommand{ \schionepm }{\mbox{$\tilde{\chi}_{1}^{\pm}$}}

\newcommand{ \isajet }    {{\tt ISAJET}}
\newcommand{ \pgs }    {{\tt PGS4}}
\newcommand{ \darksusy }    {{\tt DARKSUSY}}

\begin{document}

\preprint{MIFP-08-02}

\title{Determining the Dark Matter Relic Density in the
Minimal Supergravity 
Stau-Neutralino Coannihilation Region at the Large Hadron Collider}
\date{\today}

\author{Richard Arnowitt, Bhaskar Dutta, Alfredo Gurrola,
Teruki Kamon, Abram Krislock, and David Toback}

\affiliation{Department of Physics, Texas A\&M University, College Station, TX 77843-4242, USA}
\vspace{.5cm}

\begin{abstract}
We examine the stau-neutralino coannihilation (CA) mechanism
of the early universe.
We use the minimal supergravity (mSUGRA) model
and show that from measurements at the Large Hadron Collider
 one can predict the
dark matter relic density with an uncertainty of 6\% with 30 \invfb\ of data,
which is  comparable to the direct measurement by 
Wilkinson Microwave Anisotropy Probe.
This is done by measuring four mSUGRA parameters 
$m_0$, $m_{1/2}$, $A_0$ and $\tan\beta$ without requiring direct measurements
of the top squark and bottom squark masses.
We also provide precision measurements of
the gaugino, squark, and lighter stau masses
in this CA region without assuming gaugino universality. 

\end{abstract}

\pacs{11.30.Pb, 12.60.Jv, 14.80.Ly}
\maketitle


One of the important aspects of supersymmetry (SUSY), 
particularly when it is combined with supergravity grand unification 
(SUGRA GUT)~\cite{sugra1,sugra2}, is that it resolves a number 
of the problems inherent in the standard model (SM).
Aside from solving the gauge hierarchy problem and 
predicting grand unification at the GUT scale $M_{\rm G} \sim 10^{16}$ GeV,
subsequently verified at LEP~\cite{LEP}, 
SUGRA GUT allows for the spontaneous breaking of
SUGRA at the $M_{\rm G}$ scale in a hidden sector, 
leading to an array of soft breaking masses. 
The renormalization group equations then show
that this breaking of SUGRA leads naturally to the breaking of 
$SU(2) \times U(1)$ of the SM at the electroweak scale,
with SUSY breaking masses around a TeV for most of 
the SUSY parameter space.

An additional feature of SUSY is that models with $R$-parity invariance
 give rise to a cold dark matter (CDM) candidate~\cite{gw}, 
which is generally the lightest neutralino ($\schionezero$). 
The CERN Large Hadron Collider (LHC) should be
able to produce the $\schionezero$, and study its properties.
Direct detection experiments for  Milky Way DM
would allow for a determination of the DM mass and its nuclear cross section.
If these are in agreement with the
LHC determination of the $\schionezero$ properties, it would help confirm the important
point that the Milky Way DM was indeed the $\schionezero$.
However, this would not explicitly verify that the $\schionezero$ was the DM
relic particle produced during the Big Bang. To do this,  one would need
to deduce the relic density $\Omega_{\schionezero} h^2$
and compare with $\Omega_{\rm CDM} h^2$
as measured astronomically by 
Wilkinson Microwave Anisotropy Probe (WMAP)~\cite{WMAP}.

In this Letter we describe a series of measurements 
in the  stau-neutralino ($\stauone$-$\schionezero$)
coannihilation (CA) region where, in the early universe,
 the $\stauone$ and the $\schionezero$ annihilate
together into SM particles, determining the relic DM abundance observed today. 
We show how to  measure the sparticle masses, 
confirm we are in the CA region, 
measure the SUSY parameters,
and establish a prediction of $\Omega_{\schionezero} h^2$.

To carry out this analysis it is necessary to assume a model
that encompasses both LHC phenomena and early universe physics. 
Since the analysis is new and quite complicated
we consider the simplest SUGRA model (minimal SUGRA or mSUGRA) \cite{sugra1} 
with universal soft breaking masses. 
However, we show below that it is possible to test experimentally gaugino universality;
other non-universality models will be considered elsewhere.
The mSUGRA model  depends on  one sign and
four  parameters: $\mzero$ (universal sfermion mass), 
$\mhalf$ (universal
gaugino mass), $\azero$ (universal soft breaking trilinear coupling constant),
$\tan\beta$ (the ratio of vacuum expectation values of two Higgs doublets),
and the sign of $\mu$ (the bilinear
 Higgs coupling constant). 
 After we include all experimental constraints~\cite{LEP}, 
the allowed mSUGRA parameter space with $\mu > 0$
(as preferred by $b\rightarrow s\gamma$ 
and the muon $g-2$ \cite{BNLg-2}) has three distinct regions picked out by the
CDM constraints \cite{darkrv}: 
(i) the CA region where both $m_0$ and  $m_{1/2}$ can be small, 
(ii) the focus-point region where the $\schionezero$ 
has a large Higgsino component and $m_0$ is very large
but $m_{1/2}$ is small, and 
(iii) the funnel region where 
both $m_0$ and $m_{1/2}$ are large and 
the neutralinos can annihilate through 
 heavy Higgs bosons ($2 M_{\schionezero} \simeq M_{A_{0},H_{0}}$). 
We consider here the
CA region with $\mu > 0$.
This region is generic for a wide class
of SUGRA GUT models (with or without gaugino universality). 
If the muon $g-2$ anomaly  maintains, then the focus-point 
and funnel regions are essentially eliminated. 

The CA region has a
striking characteristic of the $\stauone$ and $\schionezero$ 
being nearly degenerate $i.e.$, 
$\dM \equiv M_{\stauone} - M_{\schionezero} \sim$ (5-15) GeV. 
Thus, the $\schitwozero \rightarrow 
\tau \stauone \rightarrow \tau\tau \schionezero$ decays are  dominant and 
 the branching ratio for
$\schitwozero \rightarrow \ell \sleptonR \rightarrow \ell\ell \schionezero$ is essentially zero ($\ell$= $e$ or $\mu$, and
\sleptonR\ is the lighter selectron or smuon). 
The existence of this near degeneracy would be
a strong indication that we are in the CA region.

In order to determine $\Omega_{\schionezero} h^2$
one must know all the mSUGRA parameters.  
In a previous study~\cite{polsello}, 
\mhalf, \mzero, \azero, and $\tan\beta$
were determined in the bulk region 
assuming that it is possible to measure the 
gluino (\gluino), squark (\squark),
lighter bottom squark (sbottom or \sbottomone), \sleptonR,
\schitwozero, and \schionezero\ masses in
$\gluino \rightarrow b \sbottomone$ and
$\squarkL \rightarrow q \schitwozero \rightarrow 
q \ell \sleptonR \rightarrow q \ell \ell \schionezero$ decays,
using   ``end-point'' techniques~\cite{hinch1}.
The determination of $M_{\tilde b_1}$  
is very difficult if both
$\gluino \rightarrow t \stopone$ 
(here \stopone\ is the lighter top squark or stop) and 
$\gluino \rightarrow b \sbottomone$ can occur~\cite{gmo} and the methodology of disentangling this background is not known yet.
Also these techniques cannot be utilized for the CA case 
because the 
$\schitwozero \rightarrow \ell \sleptonR \rightarrow \ell \ell \schionezero$ decay
is essentially absent. 

While the CA region is particularly challenging, 
we show  that it is indeed possible to
determine all four parameters accurately from measurements
at the LHC. 
 It has been recently shown~\cite{LHCtwotau,LHCthreetau} that
the CA region can be established and that
a measurement of \dM\ can be made 
(provided the $\tau$ identification
can be done for visible $\tau$ $\ptvis > 20\ \gev$)
assuming $A_0$ and $\tan\beta$ are known.
The small \dM\ value is experimentally characterized by a low energy $\tau$
from a  $\stauone \rightarrow \tau \schionezero$ decay.
With the addition of 
some new datasets and variables, in particular with final state $b$-quark jets, 
we show that we can
(a) measure the
 \gluino, \squarkL, 
\schitwozero, \schionezero, and \stauone\ masses 
in the case of $M_{\squark} \simeq M_{\gluino} \gg M_{\schitwozero, \schionepm}$
without the mSUGRA assumption,
(b) determine the mSUGRA parameters,
 and 
(c) predict $\Omega_{\schionezero} h^2$,
which can be compared with
the astronomical determination of
$\Omega_{\rm CDM} h^2$. 
Of particular note, our method
effectively obviates the need to separate
the final states arising from
the third generation sparticles, such as 
stops (\stopone, \stoptwo),
sbottoms (\sbottomone, \sbottomtwo), and
staus (\stauone, \stautwo).
The procedure of extracting  the model parameters
is general and can be applied to other regions of the
parameter space or to more general SUGRA
models. 

We select an mSUGRA reference point, shown in Table~\ref{tab:SUSYmass},
where $\Omega_{\schionezero} h^2$ = 0.10 and \dM\ = 10.6 GeV. 
The total production cross section at the LHC is 9.1 pb where the
 $\gluino\squark$  production
has the largest contribution. 
Events are generated 
using \isajet~\cite{isajet}, followed by the \pgs\ detector simulation~\cite{pgs}. 
We analyze three samples 
with the final state of large transverse missing energy (\met)
along with jets ($j$'s), $\tau$'s, and $b$'s: 
(i) 2$\tau$ + 2$j$ + \met, 
(ii) 4$j$ + \met, and 
(iii) 1$b$ + 3$j$ + \met.
The kinematics in both  2$\tau$ + 2$j$ + \met\ 
and the 1$b$ + 3$j$ + \met\ samples depend on all four mSUGRA parameters, 
while the 4$j$ + \met\ sample is mostly sensitive to $m_0$ and $m_{1/2}$. 


\begin{table}[b]
\caption{SUSY reference point (masses in GeV) 
$m_{1/2}$ = 350~GeV, $m_0$ = 210 GeV, $\tan\beta = 40$, 
$\azero = 0$, and $\mu > 0$. }
 \label{tab:SUSYmass}
\begin{center}
\begin{tabular}{c c c c c c c c c}
\hline \hline  
$\gluino$ & 
$\begin{array}{c} \usquarkL \\ \usquarkR \end{array}$ & 
$\begin{array}{c} \stoptwo \\ \stopone \end{array}$ & 
$\begin{array}{c} \sbottomtwo \\ \sbottomone \end{array}$ & 
$\begin{array}{c} \seleL \\ \seleR \end{array}$ & 
$\begin{array}{c} \stautwo \\ \stauone \end{array}$ & 
$\begin{array}{c} \schitwozero \\ \schionezero \end{array}$ &
$\dM$ &
$\begin{array}{c} M_{\gluino}/M_{\schionezero} \\ 
M_{\gluino}/M_{\schitwozero} \end{array}$ 
\\ \hline
831&
$\begin{array}{c} 748 \\ 725 \end{array}$ & 
$\begin{array}{c} 728 \\ 561 \end{array}$ & 
$\begin{array}{c} 705 \\ 645 \end{array}$ & 
$\begin{array}{c} 319 \\ 251 \end{array}$ & 
$\begin{array}{c} 329 \\ 151.3 \end{array}$ & 
$\begin{array}{c} 260.3 \\ 140.7 \end{array}$ & 
10.6 &
$\begin{array}{c} 5.91 \\ 3.19 \end{array}$ 
\\ \hline \hline
\end{tabular}
\end{center}
\end{table}
 
The primary SM backgrounds for the 2$\tau$ + 2$j$ + \met\ final state
(and the other two samples) are 
from $t\bar{t}$, $W$+jets, $Z$+jets and QCD production. 
The sample is selected using the following  cuts~\cite{LHCtwotau}:
(a) $N_{\tau} \geq 2$  ($|\eta| < 2.5$, $\ptvis >20\ \gev$,
but $>40\ \gev$ for the leading $\tau$);
(b) $N_{j} \geq 2$  ($|\eta| < 2.5$, $\et > 100\ \gev$);
(c) $\met > 180\ \gev$ and
$\et^{j\rm{1}} + \et^{j\rm{2}}$ + \met\ $>$ 600 GeV; and
(d) veto the event if any of the two leading jets are identified 
as a $b$ jet.
In order to identify 
$\schitwozero
\rightarrow \tau\stauone
\rightarrow \tau\tau\schionezero$ decays
we categorize all pairs of $\tau$'s 
into opposite-sign (OS) and like-sign (LS) combinations,
 and then use the OS minus LS (OS$-$LS) distributions 
to effectively reduce the SM events as well as the
combinatoric SUSY backgrounds.
We reconstruct  the  decay chains of 
$\squarkL \rightarrow q \schitwozero 
\rightarrow q \tau \stauone \rightarrow q \tau\tau \schionezero$
using the following five kinematic variables:
(1) $\alpha$, the slope of the \ptvis\ distribution for the lower energy $\tau$
in the OS$-$LS di-$\tau$ pairs,  
(2) $M_{\tau\tau}^{\rm peak}$, the peak position 
of the visible di-$\tau$ invariant mass distribution,  
(3) \mjttpeak, the peak position  of the invariant $j$-$\tau$-$\tau$ mass distribution,   
and
(4,~5) \mjtpeak, the peak position of the invariant $j$-$\tau$ mass distribution 
where each $\tau$ from the OS$-$LS di-$\tau$ pair is examined separately. 
Note that we have used the peak positions 
instead of the end-points because of the $\tau$'s in the final state.

We follow the recommendation of Ref.~\cite{hinch1} for the 4$j$ + \met\ sample.
The peak value, $M_{\rm eff}^{\rm peak}$, of the variable $\meff \equiv \met\ + \sum_{{\rm 4~jets}} \et^{j}$, which 
is a function of only the $\gluino$ and $\squark$ masses,
is reconstructed for each event that passes
the following selection cuts:
(a) $N_{j} \geq 4$  ($|\eta| < 2.5$, $\et > 50\ \gev$, 
but  $> 100\ \gev$ for the leading jet);
(b) $\met > 100\ \gev$;
(c) Transverse sphericity $>$ 0.2; 
(d) Veto on all events containing an isolated electron or muon with $\pt > 15\ \gev$
and $|\eta| < 2.5$; and
(e) $\met > 0.2 \meff$.
Again we require that none of these jets be identified as  a $b$ jet.

Similar cuts are used to make 
the 1$b$ + 3$j$  + \met\ sample. 
We introduce a new variable,
\meffbpeak, similar to \meffpeak, but requiring  
that the leading jet be from a $b$ quark.


\begin{figure}
\centering
\includegraphics[width=.35\textwidth,angle=-90]{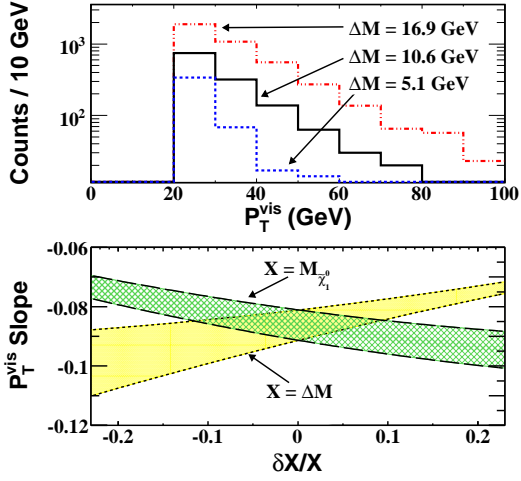}
\caption{[top] The $\ptvis$ distribution  of the lower-energy $\tau$'s
using the OS$-$LS technique in the three samples (arbitrary luminosity) of SUSY events 
with \dM\ = 5.1, 10.6 and 16.9 GeV, 
where only the \stauone\ mass is changed at our reference point.
[bottom] The $\ptvis$ slope (as $\alpha$ in the text)
 as a function of the relative change of 
 $M_{\tilde\tau_1}$ (therefore $\dM$) or $M_{\schionezero}$
from its reference value with the other SUSY masses fixed.
 The bands correspond to statistical uncertainties with 10 \invfb.
}
\label{fig:visTauPt}
\end{figure}


The measurement of a small value of $\alpha$  from the 2$\tau$ + 2$j$ + \met\ sample indicates low energy $\tau$'s in the final state 
(thus \dM\ is small) and 
 provides  a smoking-gun signal for the CA region.
In Fig.~\ref{fig:visTauPt}, we show the \ptvis\ distributions obtained
by the OS$-$LS technique for various \dM\ values.
Note that $\alpha$ only depends on 
$M_{\stauone}$ and $M_{\schionezero}$ (see Fig.~\ref{fig:visTauPt}). 

\begin{figure}[b]
\centering
\includegraphics[width=.35\textwidth,angle=-90]{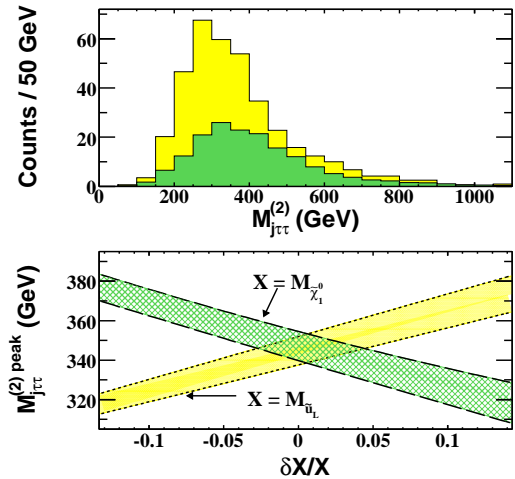}
\caption{[top] The $M_{j\tau\tau}^{(2)}$ distributions using the OS$-$LS technique
for SUSY events at our reference point, but with
$M_{\squarkL}$ = 660 GeV (yellow or light gray histogram) 
and 840 GeV (green or dark gray histogram), where 748 GeV is  
our reference point;
[bottom] The peak position of the mass distribution
 as a function of $M_{\schionezero}$ or  $M_{\squarkL}$.
 The bands correspond to statistical uncertainties with 10 \invfb.
}
\label{fig:Mjtautau}
\end{figure}

To get a set of measurements of the sparticle masses
we use the remaining variables from 
the 2$\tau$ + 2$j$ + \met\ and 4$j$ + \met\ samples. 
The variables \mjttpeak\ and \mjtpeak\ probe the 
$\squarkL \rightarrow q \schitwozero 
\rightarrow q \tau\stauone 
\rightarrow q \tau\tau\schionezero$ decay chains. 
To help identify these  chains we additionally require
 OS$-$LS di-$\tau$ pairs with 
$M_{\tau\tau} < M_{\tau\tau}^{\mbox{\rm end-point}}$
and construct $M_{j\tau\tau}$ for every jet with
$\et > 100\ \gev$ in the event.
With three jets, there are
three masses: $M_{j\tau\tau}^{(1)}$,
$M_{j\tau\tau}^{(2)}$, and
$M_{j\tau\tau}^{(3)}$, in  decreasing order.
We choose 
 $M_{j\tau\tau}^{(2)}$ for this analysis~\cite{hinch1}.
Figure~\ref{fig:Mjtautau}  shows
the $M_{j\tau\tau}^{(2)}$ distributions for two different \squarkL\ masses,
and  $M_{j\tau\tau}^{(2)\rm peak}$ as a function of
$M_{\squarkL}$ and $M_{\schionezero}$, 
 keeping \dM\ constant. 
Similarly, one can show that 
the $M_{j\tau}^{(2)\rm peak}$ value depends on the \squarkL, 
\schitwozero, \stauone\  and  \schionezero\  masses.
The value of \meffpeak, 
extracted from 4$j$ + \met\ sample,
has been shown to be  a function of  
only the \squarkL\ and \gluino\ masses.


The determination of the sparticle masses 
is done by inverting the six functional relationships between the variables
and the sparticle masses
to simultaneously solve 
for the \gluino, $\tilde\chi^0_{1,2}$, \stauone, and
average \squarkL\ masses and their uncertainties. 
The six parametrized functions are:
 \mtautaupeak\ = $f_1$($M_{\schitwozero}$, $M_{\schionezero}$, $\dM$),
 $\alpha$ = $f_2$($M_{\schionezero}$, $\dM$),
 $M_{j\tau\tau}^{(2)\rm peak}$ = 
 $f_3$($M_{\squarkL}$, $M_{\schitwozero}$, 
$M_{\schionezero}$),
$M_{j\tau(1,2)}^{(2)\rm peak}$ = $f_{4,5}$($M_{\squarkL}$,
 $M_{\schitwozero}$, $M_{\schionezero}$, $\dM$),
 and 
 \meffpeak\ = $f_6$($M_{\squarkL}$, $M_{\gluino}$).
With 10 \invfb\ of data, 
we obtain (in GeV)
$M_{\gluino} = 831 \pm 28$,
$M_{\schitwozero} = 260 \pm 15$, 
$M_{\schionezero} = 141 \pm 19$,
$\dM = 10.6 \pm 2.0$, and
$M_{\squarkL} = 748 \pm 25$~\cite{uncertainties}.
The accurate determination of $\Delta M$  
would also confirm  that we are in the CA region.
We also test the universality of the 
gaugino masses at the GUT scale.
We measure $M_{\gluino}/M_{\schionezero}=5.9 \pm 0.8$ and $M_{\gluino}/M_{\schitwozero}=3.1 \pm 0.2$,  
validating the universality relations to $14 \%$ and $6 \%$, repectively. 
This non-trivial  determination of 
the additional gaugino masses along 
with the mSUGRA parameters require all six observables. 
The formalism developed here can work for  other model 
with similar two-body decay processes.


\begin{figure}[b]
\centering
\includegraphics[width=.35\textwidth,angle=-90]{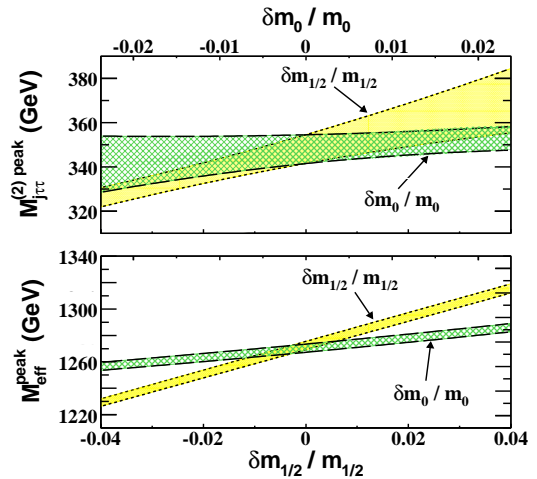}
\caption{The dependence of $M_{j\tau\tau}^{(2)\rm peak}$ (top)
and \meffpeak\ (bottom) as a function of $m_{1/2}$ and $m_{0}$.
 The bands correspond to statistical uncertainties with 10 \invfb.
}
\label{fig:Mjtautau_mSUGRA}
\end{figure}

\begin{figure}
\centering
\includegraphics[width=.35\textwidth,angle=-90]{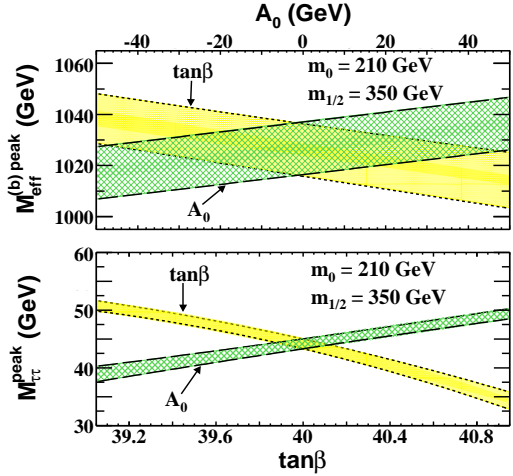}
\caption{The dependence of  $\meffbpeak$ (top)  and
\mtautaupeak\ (bottom) as a function of $\tan\beta$ and $A_{0}$.
 The bands correspond to statistical uncertainties with 10 \invfb.
}
\label{fig:Meffb}
\end{figure}

Since our primary goal is to determine 
$\Omega_{\schionezero} h^{2}$ in the mSUGRA model 
we next determine $m_0$, $m_{1/2}$, \azero\ and $\tan\beta$. 
$M_{\rm eff}^{\rm peak}$ and $M_{j\tau\tau}^{(2)\rm peak}$
are insensitive to \azero\ and $\tan\beta$, and
provide a direct handle on \mzero\ and \mhalf\
because they depend only on 
the \squarkL\ (first two generations), 
\gluino, \schitwozero\ and \schionezero\ masses
(see Fig.~\ref{fig:Mjtautau_mSUGRA}).  
On the other hand, \mtautaupeak\ and  \meffbpeak\  provide
a direct handle on \azero\ and $\tan\beta$.
\mtautaupeak\ depends on $M_{\tilde \tau_1}$;
\meffbpeak\ depends on $M_{\tilde t_1}$ and $M_{\tilde b_1}$,
since both the \stopone\ and \sbottomone\ decays always 
produce at least one $b$ jet in the final state.
Figure~\ref{fig:Meffb} shows
the values of  \mtautaupeak\ and \meffbpeak\  
as functions of \azero\ and $\tan\beta$ since the off-diagonal elements of  \stopone\ and  \sbottomone/\stauone\ mass matrices depend on
 $M_t(A_t+\mu \cot\beta)$ and  $M_{b/\tau}(A_{b/\tau}+\mu\tan\beta)$, respectively.
Combining these four measurements and inverting, 
we find 
$m_0 = 210 \pm 4\ \gev$, $m_{1/2} = 350 \pm 4\ \gev$,
$A_0 = 0 \pm 16\ \gev$, and $\tan\beta = 40 \pm 1$
with 10 \invfb\ of data~\cite{uncertainties}.


\begin{figure}[b]
\centering
\includegraphics[width=.23\textwidth,angle=-90]{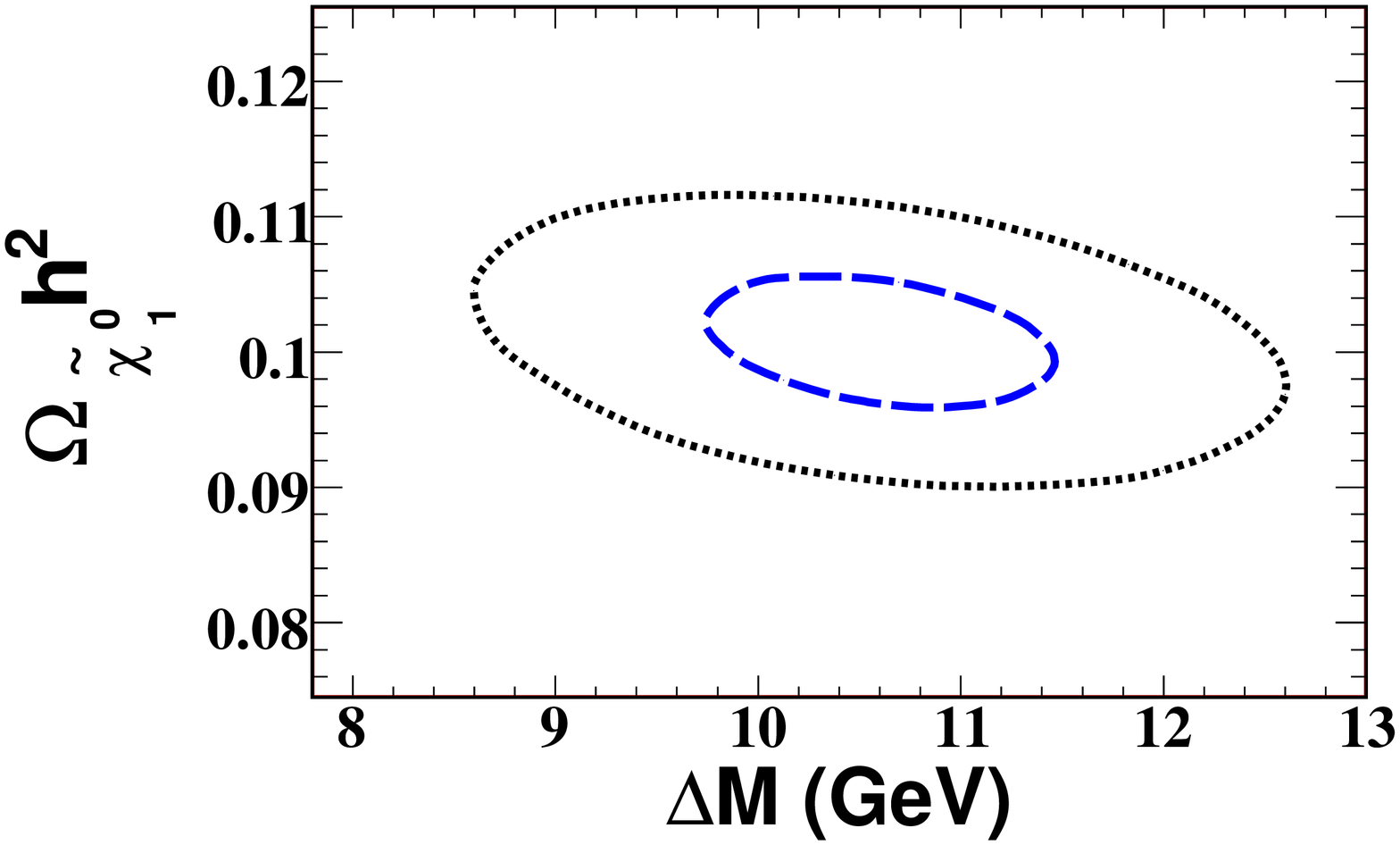}
\caption{Contour plot of the 1$\sigma$ uncertainty 
in the $\Omega_{\schionezero} h^2$-$\dM$ plane with
10  and 50 \invfb\ (outer and inner ellipses).}
\label{fig:Omegah2}
\end{figure}

After measuring the mSUGRA variables 
we  calculate $\Omega_{\schionezero} h^2$
using \darksusy~\cite{darksusy}. 
The calculation also involves 
the $\schionezero$ mixing matrix which we have determined in the mSUGRA case. 
In the CA region, $\Omega_{\schionezero} h^2$ depends crucially on $\dM$ 
due to the Boltzmann suppression factor $e^{- \Delta M/k_{B}T}$
in the relic density formula~\cite{gs}. 
Figure~\ref{fig:Omegah2} shows contour plots of 
the 1$\sigma$ uncertainty in  the $\Omega_{\schionezero} h^2$-$\dM$ plane.
The uncertainty on $\Omega_{\schionezero} h^2$ is 
11~(4.8)\% at 10~(50) \invfb~\cite{uncertainties}.
Note that it is 6.2\% at 30~\invfb~\cite{rel},
comparable to that of the WMAP measurement~\cite{WMAP}. 


In conclusion, we have established a technique for
a  precision  measurement of $\Omega_{\schionezero} h^2$
at the LHC in the \stauone-\schionezero\ CA region
of the mSUGRA model.
This is done using only the model parameters,
determined by the  kinematical analyses 
of 3 samples of  \met\ + $j$'s (+ $\tau$'s) events with and without $b$ jets.  
The accuracy of the $\Omega_{\schionezero} h^2$ calculation at 30 \invfb\
of data
is expected to be 
comparable to that of $\Omega_{\rm CDM} h^2$ by WMAP.  This approach 
will allow us to determine the relic
abundance at the LHC for any model where the CA is 
dominant in the early universe.
Thus, it is possible to confirm 
that the DM we observe today were $\tilde\chi^0_1$'s created in the early universe.

We would like to thank F.E. Paige,
M.M. Nojiri, G. Polesello, and D.R. Tovey for useful discussions.
This work was supported in part by a DOE grant DE-FG02-95ER40917
and NSF grant DMS 0216275. A.G. is supported by DOEd GAANN.


\begin{thebibliography}{99}

\bibitem{sugra1}
A.H. Chamseddine, R. Arnowitt, and P. Nath,   
  Phys. Rev. Lett.  {\bf 49}, 970 (1982).


\bibitem{sugra2}
L. Hall, J. Lykken, and S. Weinberg,  
  Phys. Rev.  D{\bf 27}, 2359 (1983);
P. Nath, R. Arnowitt, and A.H. Chamseddine, 
  Nucl. Phys.  B{\bf 227}, 121 (1983).
For a review, see P. Nilles, Phys. Rep. {\bf 100}, 1 (1984).


\bibitem{LEP}
Particle Data Group,  S.~Eidelman \etal,
  Phys. Lett. B{\bf 592}, 1 (2004).
\bibitem{gw}
  H.~Goldberg,
  Phys.\ Rev.\ Lett.\  {\bf 50}, 1419 (1983).


\bibitem{WMAP}
WMAP Collaboration, D.N. Spergel {\it et al.}, Astrophys. J. Suppl. {\bf 148} (2003) 175.

\bibitem{BNLg-2}
Muon $g-2$ Collaboration, G.~W.~Bennett {\it et al.},
  Phys. Rev. Lett. {\bf 92}, 161802 (2004);
S.~Eidelman,
  Acta Phys. Polon.  B{\bf 38}, 3015 (2007).

\bibitem{darkrv}
J. Ellis \etal,
  Phys. Lett.  B{\bf 565}, 176 (2003);
R. Arnowitt,
B. Dutta, and B. Hu,  
  arXiv:hep-ph/0310103;
H. Baer \etal,
  J. High Energy Phys. {\bf 06} (2003) 054;
B. Lahanas and D.V. Nanopoulos, Phys. Lett. B{\bf 568}, 55 (2003); 
U. Chattopadhyay, A. Corsetti, and P. Nath, 
  Phys. Rev.  D{\bf 68}, 035005 (2003);
E. Baltz and P. Gondolo, 
  J. High Energy Phys. {\bf 10} (2004) 052.

\bibitem{polsello} 
G.~Polesello and D.~R.~Tovey,
  J. High Energy Phys. {\bf 05} (2004) 071;
  M.~M.~Nojiri, G.~Polesello, and D.~R.~Tovey,
  J. High Energy Phys. {\bf 03} (2006) 063.

\bibitem{hinch1}
I. Hinchliffe \etal, 
  Phys. Rev.  D{\bf 55}, 5520 (1997);
I. Hinchliffe and F.E. Paige,
  Phys. Rev.  D{\bf 61}, 095011 (2000).
 

\bibitem{gmo}
B.K. Gjelsten, D.J. Miller, and P. Osland,
J. High Energy Phys. {\bf 12} (2004) 003. 

\bibitem{LHCtwotau}
R. Arnowitt \etal, 
  Phys.\ Lett.\  B{\bf 639}, 46 (2006).

\bibitem{LHCthreetau}
R. Arnowitt \etal, 
  Phys.\ Lett.\  B{\bf 649}, 73 (2007).

\bibitem{isajet}
F.E.~Paige \etal,
  arXiv:hep-ph/0312045.
 We use ISAJET version 7.64 with  TAUOLA.

\bibitem{pgs}
PGS is a parameterized detector simulator.
We use version 4 
(\url{http://www.physics.ucdavis.edu/~conway/research/software/pgs/pgs4-general.htm})
in the CMS detector configuration.
We assume the $\tau$ identification efficiency
with $\ptvis > 20\ \gev$ is 50\%, while
the probability for a jet being mis-identified 
as a $\tau$ is  1\%. 


\bibitem{uncertainties}
All uncertainties are statistical. 
The systematic uncertainties 
will be evaluated correctly once the LHC starts.


\bibitem{darksusy}
P.~Gondolo \etal,
  arXiv:astro-ph/0211238.

\bibitem{gs}
  K.~Griest and D.~Seckel,
  Phys. Rev.  D{\bf 43}, 3191 (1991).

\bibitem{rel}The $\Omega_{\schionezero} h^2$ is calculated at  tree-level.
Since the dominant diagrams involve  only non-colored particles  the one loop QCD corrections are also very small.


\end{thebibliography}

\end{document}